# Design and construction of an optical test bed for LISA imaging systems and tilt-to-length coupling

M Chwalla[1], K Danzmann[2], G Fernández Barranco[2], E Fitzsimons[1,4], O Gerberding[2], G Heinzel[2], C J Killow[3], M Lieser[2], M Perreur-Lloyd[3], D I Robertson[3], S Schuster[2], T S Schwarze[2], M Tröbs[2], H Ward[3] and M Zwetz[2]

[1] Airbus DS GmbH, Claude-Dornier-Straße, 88090 Immenstaad, Germany
[2] Max Planck Institute for Gravitational Physics (Albert Einstein Institute) and Institute for Gravitational Physics of the Leibniz Universität Hannover, Callinstr.38, 30167 Hannover, Germany
[3] SUPA, Institute for Gravitational Research, University of Glasgow, Glasgow G12 8QQ, Scotland, UK

E-mail: michael.troebs@aei.mpg.de

## Abstract

The laser interferometer space antenna (LISA) is a future space-based interferometric gravitational-wave detector consisting of three spacecraft in a triangular configuration. The interferometric measurements of path length changes between satellites will be performed on optical benches in the satellites. Angular misalignments of the interfering beams couple into the length measurement and represent a significant noise source. Imaging systems will be used to reduce this tilt-to-length coupling.

We designed and constructed an optical test bed to experimentally investigate tilt-to-length coupling. It consists of two separate structures, a minimal optical bench and a telescope simulator. The minimal optical bench comprises the science interferometer where the local laser is interfered with light from a remote spacecraft. In our experiment, a simulated version of this received beam is generated on the telescope simulator. The telescope simulator provides a tilting beam, a reference interferometer and an additional static beam as a phase reference. The tilting beam can either be a flat-top beam or a Gaussian beam. We avoid tilt-to-length coupling in the reference interferometer by using a small photo diode placed at an image of the beam rotation point. We show that the test bed is operational with an initial measurement of tilt-to-length coupling without imaging systems.

[4] Present address: UK Astronomy Technology Centre, Royal Observatory Edinburgh, Blackford Hill, Edinburgh EH9 3HJ, UK

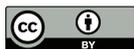

Furthermore, we show the design of two different imaging systems whose performance will be investigated in future experiments.

(Some figures may appear in colour only in the online journal)

## 1. Introduction

The laser interferometer space antenna (LISA) is a future space-based gravitational wave detector consisting of three satellites in heliocentric orbits [1, 2].

Laser-links between the spacecraft are used for measuring the tiny distance fluctuations between free-floating test masses inside the satellites caused by gravitational waves. Telescopes are used for sending and receiving light between spacecraft. The interferometric path length measurements are split in different parts and each satellite has optical benches with several interferometers. To detect gravitational waves these individual measurements are combined to form a Michelson-like interferometer. The freely-floating test masses and local interferometry have recently been demonstrated on the LISA Pathfinder (LPF) spacecraft [3].

Each optical bench (OB) is situated between a telescope and a test mass. The interface between the telescope and the OB is the receive (Rx) clip which is the entrance pupil of the telescope. The received beam will tilt due to spacecraft jitter and breathing of the triangular constellation and it will rotate around the Rx clip on the OB.

Tilt-to-length coupling is currently the second largest entry in the LISA metrology error budget (after shot noise) [4]. The requirement on the imaging systems is that they should suppress the coupling from tilt to length to less than $\pm 25$ $\mu$m rad$^{-1}$ for beam angles on the optical bench within $\pm 300$ $\mu$rad [5].

Recently, a proof-of-principle experiment was reported on that demonstrated a reduction in tilt-to-length coupling by using a two-lens imaging system [6]. This investigation was not fully representative of LISA because it did not include effects of higher-order modes and different beam parameters of the interfering beams.

We describe the design and the build of a test bed to experimentally investigate tilt-to-length coupling in a way as representative as possible of LISA. Different beam parameters of the interfering beams and higher-order modes are present. Additionally, the test bed has been designed to offer a versatile platform for testing of other aspects of LISA metrology—e.g. imaging systems, alternative photoreceivers or a mechanism to compensate linear tilt-to-length coupling called active aperture—which may be desirable in the future.

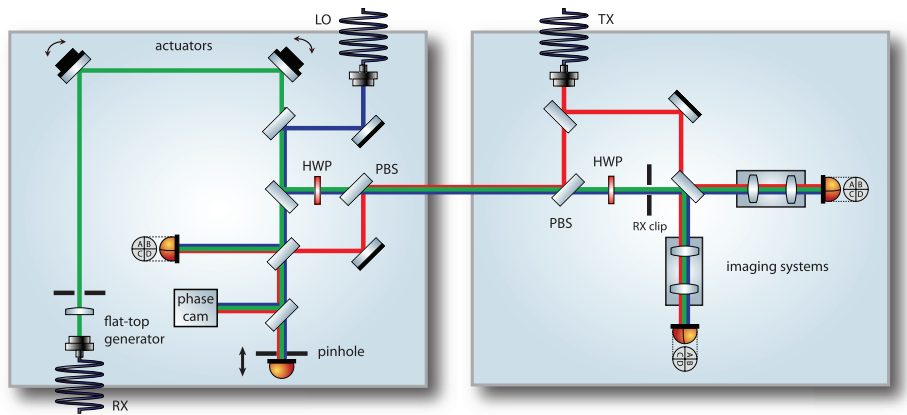

**Figure 1.** Schematic of the test bed concept. The telescope simulator (left) and optical bench (right) are shown with the key components to illustrate the measurement concept. The Rx beam is shown in green, the LO beam in blue and the Tx beam in red. The Rx beam is tilted around the middle of the Rx clip with the two actuators and the pinhole photodiode is placed at the same optical distance as the Rx clip. The beams between the two baseplates have a different polarization and are separated by PBSs. On the optical bench the imaging systems are placed in the science interferometer in front of the photodiodes in both output ports.

## 2. Optical test bed system design

The primary purpose of the test bed, described below, is to provide a facility in which we can demonstrate, in a representative way, that suitably designed imaging systems can be used to suppress tilt-to-length coupling for LISA. Specifically, we aim to be able to:

- Verify, by measurement of the tilt-to-length coupling, that imaging systems can reduce tilt-to-length coupling to the required level
- Investigate the effect of imaging system alignment on the measured coupling
- Investigate the effect of photoreceiver misalignment on the measured coupling
- Test different designs of imaging systems (classical pupil imaging and non-classical)

To facilitate these measurements, we require a test set-up which has the following minimum set of features:

- A measurement interferometer, with space for an interchangeable imaging system
- A reference interferometer against which to make the measurement
- A static reference beam (called Tx beam)
- A measurement beam (called Rx beam) which can be tilted about the entrance pupil of the imaging system under test

The difficulty in making such a measurement is in the separation of the effects; it must be possible to disentangle any tilt-to-length coupling generated by the test equipment from the coupling we wish to measure—that remaining with the imaging systems in the science interferometer.

Figure 1 shows a schematic of the test bed concept. It consists of two parts: the optical bench and the telescope simulator (TS). The first is a representation of the main optical instrument on board the LISA satellites, the latter is a tool to characterize the optical bench. Both parts in combination form the optical test bed.

The telescope simulator produces a tiltable flat-top beam (Rx) that represents the beam received on a LISA satellite from the remote satellite. This beam is tilted around the Rx clip on the optical bench in the same way as a beam received by a telescope on board a LISA satellite. On the optical bench the Rx beam is interfered with a part of the local transmit (Tx) beam in the science interferometer where both beams are detected by a photodiode. Additionally, the TS delivers a static beam (LO) that is used as alignment and phase reference. That LO beam will not be present in LISA but is an auxiliary tool for these measurements. The imaging system to be tested is the device under test (DUT) that is placed in front of the photodiode. A fraction of the Tx beam is split off before the interference with the Rx beam and is sent to the telescope simulator. Polarising beam splitters on optical bench and telescope simulator are used to separate Tx and Rx beams.

Both interferometers—reference and measurement—are illuminated with all three beams (LO, Tx and Rx). This leads to three interference signals: A: Rx–Tx, B: LO–Tx, C: Rx–LO. Tilt-to-length coupling is evaluated by tilting the Rx beam and measuring the length change in signal A (Rx–Tx) between reference and measurement interferometer. At the reference interferometer, the signal between LO and Rx is kept constant on the reference SEPD photodiode by stabilizing the length of the Rx path to that of the LO path (see figure 5). This removes any path length errors introduced by the actuators. Additionally, the length of the Tx path is adjusted to that of the LO path (also shown in figure 5). In this way, the static, stable LO beam becomes the phase reference for the experiment. Crucial to this approach is that the distance from the actuators to the reference detector must be identical (with mm accuracy) to the distance to the Rx clip—not the optical pathlength $\Sigma_i n_i \cdot d_i$ but $\Sigma_i d_i/n_i$, where $n_i$ is the refractive index and $d_i$ the geometrical length of segment $i$. The latter quantity is relevant for the mode propagation of Gaussian laser beams and higher orders. This will ensure that we can create a tilting Rx beam without tilt-to-length coupling at the location of the Rx clip—which is the entrance pupil for the imaging system under test. The exit pupil is on the photodiodes in the measurement interferometer. Thus, the tilt-to-length coupling measured in the measurement interferometer is that induced by any imperfect imaging of the device under test.

## 3. Optical bench design

The optical bench (OB) components are laid out on a Zerodur® glass-ceramic baseplate of diameter 580 mm, thickness 80 mm and mass 55 kg, obtained for an earlier design of the experiment that would have required a greater density of optical components [7, 8]. The baseplate thickness had been chosen to minimize bending, to ensure satisfactory positional stability of all mounted components throughout the build and also measurement accuracy during testing. Mounting of the baseplate is via three 20 mm diameter chrome steel ball-bearings that are glued into hemispherical cutaways in the bottom surface of the substrate. These steel ball-bearings are in contact with mating features in a metal kinematic interface plate. The optical layout is shown in figure 2.

The key features—the interferometric detector assemblies (IDAs), Rx clip, Tx fibre injector optical sub-assembly (FIOS), TS Interface, TS outline and the beam combiner of the measurement interferometer (BS21) are indicated. The red beam is the locally generated, static Tx beam. The green beam is the Rx beam produced on the telescope simulator. Also present is the blue, static, auxiliary LO beam. The LO beam facilitates alignment and follows the same path on the OB as the Rx beam. The optical layout comprises a stable ∼2 mm diameter Tx beam launched from a monolithic fused silica FIOS (see section 7.1). Collimated 1064 nm light from this injector is combined with the Rx beam of the telescope simulator at the beam combiner of the measurement interferometer, BS21, then directed through each imaging system and detected at their respective photodiodes. The OB design incorporates an on-board alignment aid consisting of two spatially separated QPDs, much like a calibrated quadrant photodiode pair (CQP) [9]. These were fixed and aligned during manufacture such that any beam from the telescope simulator intersecting at the center of both QPDs would then optimally complete the measurement interferometer. This was extremely effective in enabling the precision bonding of BS21.

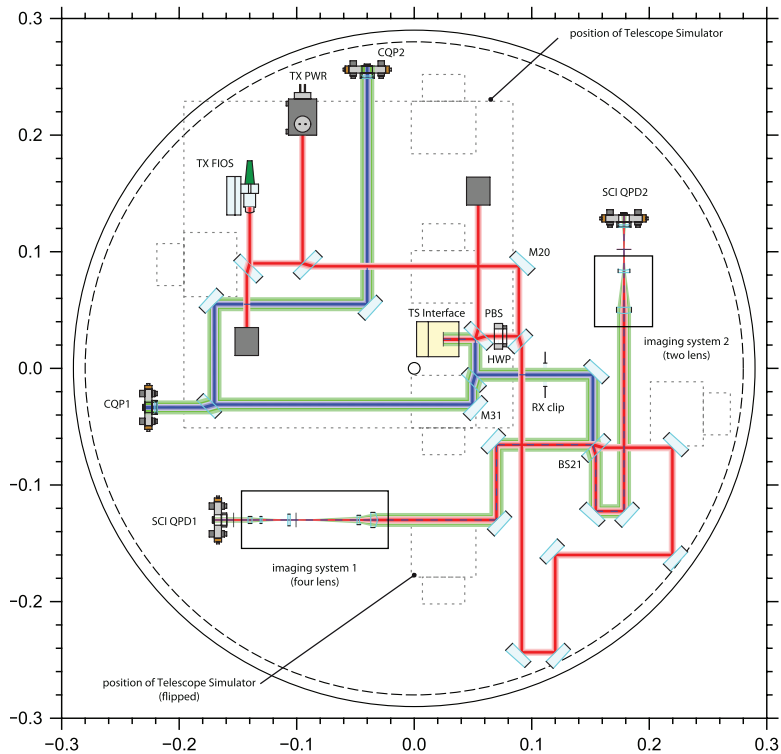

**Figure 2.** Optical layout of the optical bench (OB) and labeling of the key components. The imaging systems in front of the SCI (science interferometer) quadrant photodiodes (QPDs) are on separate baseplates and can be exchanged. Here, one four-lens and one two-lens imaging system are shown. The Rx (green) and the LO (blue) beam are produced on the telescope simulator and interfered with the Tx (red) beam from the TX FIOS. The two CQP (calibrated quadrant photodiode pair) photodiodes are used for the alignment. The dashed outline is indicating the position of the telescope simulator in the nominal and the flipped position. The scale is in meters.

Optical components such as mirrors and beamsplitters are manufactured from Corning 7980 HPFS Class 0 fused silica with tight control over the surface parallelism of the optical surfaces—2 arc seconds—and perpendicularity of the bottom surface—2 arc seconds—to facilitate the precision hydroxide-catalysis bonding [10]. The optical and bonding surfaces were polished with a resulting flatness of lambda/12 (measured at 633 nm) over 90% of the surface. These component specifications have successful heritage from the LISA Pathfinder optical bench interferometer [11]. In a change from LPF, mirrors have horizontally angled rear surfaces to prevent reflection of beams from the back surface returning along the path from where they came. The component labeled 'TS Interface' is another difference to LPF.

This 45° 'periscope' mirror directs the Tx and Rx beams to and from the telescope onboard a LISA satellite. In our setup, the telescope is replaced by the telescope simulator.

Opto-mechanical components used on the baseplate and in the telescope simulator design such as waveplate holders, photodiode mounts and beam dumps are all designed with thermal stability and thermal-mechanical stress in mind. The photodiode mounts used for the science interferometer and the on-board CQP, glued directly to the Zerodur® optical baseplate, use a combination of titanium and aluminum in their construction. This compensates for any thermal expansion and ensures that the centre position of the mounted photodiode remains stable at the $\mu$m-level. The photodiode packages were electrically and thermally isolated by mounting in MACOR ceramic and attached to the titanium mounts with two M2 screws.

The component labeled 'Rx clip' represents the interface between optical bench and telescope simulator. The Rx beam delivered to the optical bench rotates around the Rx clip. If a 2 mm diameter aperture was placed in that position, it would be the defining aperture of the system.

The power-monitor photodiode (PWR) mounts and waveplate (HWP) holders required a less stringent approach to thermal stability and so a titanium architecture was sufficient to meet the requirements. The waveplate was glued to a bespoke mounting ring in which three flexures equally spaced around the circumference of the waveplate provided a stress-minimized interface.

To address mechanical stressing of the OB baseplate through differential thermal expansion at the mechanical interface, all metallic opto-mechanical components, including the photodiode mounts, were designed with flexure 'feet'. Hysol EA9361, a high elongation, medium strength two component epoxy with relatively good outgassing qualities and with heritage from LPF [11, 12] was selected for use at the mechanical interfaces and was, at least in part, chosen for its ability to minimize thermally induced stresses transferred to the baseplate.

## 4. Telescope simulator design

The telescope simulator (TS) has a 280 mm × 280 mm Zerodur® baseplate of thickness 35 mm and mass 7 kg that is polished for bonding on both sides. It has a hole through which the periscope optics direct the beams to and from the optical bench. Figure 3 shows the schematic of the telescope simulator layout and labeling of the components. The key features—the Rx generators, the LO FIOS, the reference detectors and the actuators are highlighted, as is the outline of the TS interface mount.

Two additional beam splitters are used in the reference interferometer to create four readout ports to maximize the system flexibility. Nominally these would consist of a reference QPD, a reference single element photodiode (SEPD), a phase camera and a trigger for the phase camera. Each port is path-matched to the Rx clip. Furthermore, the TS features twin Rx-Beam sources—one Gaussian (from a FIOS) and one flat-top—both of which can be steered via the on-bench actuators.

The flat-top generator consists of a large fiber coupler producing a 9 mm-radius Gaussian beam that is clipped by an apodized aperture (see figure 4). The aperture is optimized to minimize diffraction and result in a flat phase and flat intensity over a diameter of three millimeter at the plane of the Rx clip on the optical bench.

An LO beam from a FIOS provides the required ultra-stable reference for the TS. A polarizing beam splitter and half-waveplate facilitate the required polarization multiplexing between the OB and TS.

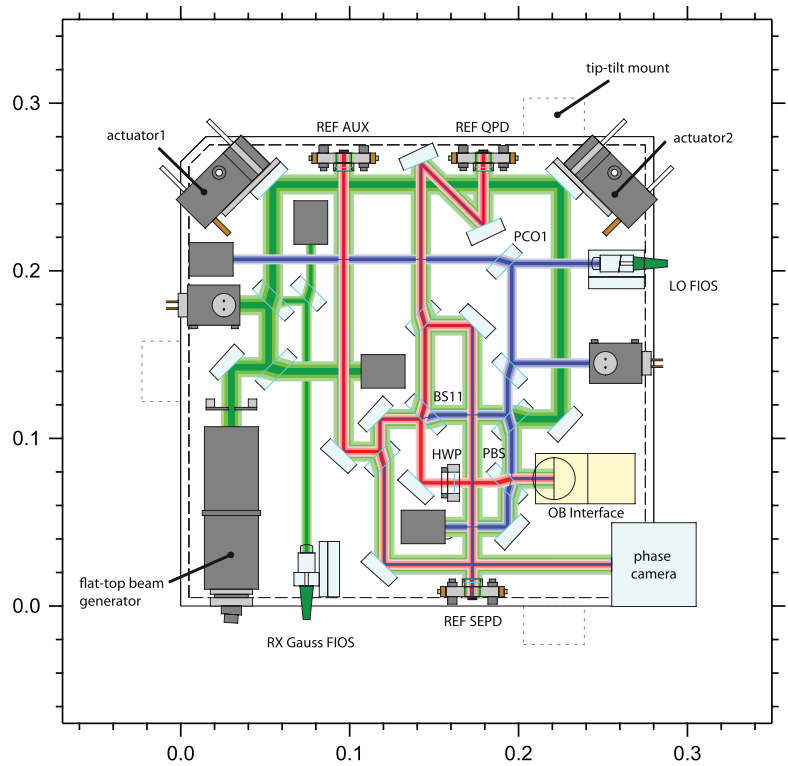

**Figure 3.** Optical layout of the telescope simulator (TS) and labeling of key components. Either the Rx flat top from the flat-top generator or the Rx Gauss from the Rx FIOS can be used. The Rx beam (green) is tilted by two piezo driven actuators and combined with the stable reference LO beam (blue). The reference interferometer has four readout ports with three different photodiodes and a phase camera. The REF SEPD is a small pinhole photodiode, the REF QPD is a quadrant photodiode and the AUX SEPD is a big single element photodiode. The dashed outlines are the positions of the feet for the tip-tilt mount to align the telescope simulator. The scale is in meters.

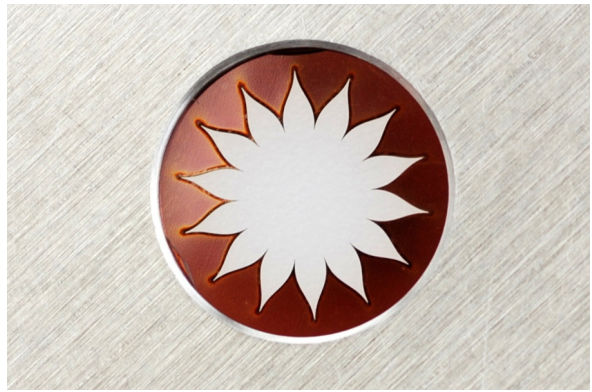

**Figure 4.** Photograph of the apodized aperture in aluminium mount. The bore in the aluminium mount has a diameter of 8 mm.

**Table 1.** Imaging system design specification for the science interferometer.

| Parameter | Specification | Comment |
| --- | --- | --- |
| Design wavelength | 1064.5 ± 2 nm | |
| Nominal distance: entrance pupil—first lens surface | 350 mm | |
| Maximum envelope: first lens surface—exit pupil | 154 mm | |
| Minimum distance: last lens surface—exit pupil | 30 mm | |
| Minimum clear aperture | 12 mm | Minimum diameter of first lens |
| Maximum lens diameter | 16 mm | Maximum physical diameter of any lens |
| Minimum lens separation | 4 mm | Minimum clear space between adjacent lenses |
| Entrance pupil diameter (minimum) | 3 mm | Rx clip diameter (2.24 mm) + alignment margin |
| Exit pupil diameter (minimum) | 1.2 mm | QPD aperture diameter (900 $\mu$m) + alignment margin |
| Magnification | 0.4 (±0.03) | relation between exit pupil diameter / entrance pupil diameter, as outlined above |
| Field of View | ±0.5 mrad | Maximum pointing range plus alignment margin |
| Maximum shift of exit pupil over field | 10 $\mu$m | phase center offset allocation × magnification |
| Maximum wavefront error within field of view | $\lambda$ peak–peak | Uncritical, should not be a design driver |
| Maximum change of wavefront over field | $\lambda/10$ rms | Difference between on-axis and tilted wavefronts |
| Minimum ghost power suppression (within exit pupil) | $10^{-6}$ | |
| Maximum scattered light in exit pupil | $10^{-4}$ | |
| Maximum tilt-to-length coupling over field | 25 $\mu$m rad$^{-1}$ | |

## 5. Imaging systems design

The main purpose of the test bed is to test the performance of imaging systems. Therefore, the properties of the imaging systems, planned for the test, are a design driver for the test bed.

The imaging systems were designed against requirements representative for LISA. A magnification factor of 0.4 was chosen to detect most power of 2 mm diameter beams on the OB on 1 mm diameter photodiodes.

A complete list of the required design specifications can be found in table 1.

### 5.1. Four-lens system optical design

The four-lens imaging system is designed using a classical optics approach (pupil plane imaging system). The resulting on-axis design uses spherical fused silica lenses with curvatures that were selected from the list of standard tools from Zeiss. The design includes a 150 $\mu$m field stop between lens two and three to block stray beams reaching the photodiode. In the coordinates of the given *x*-axis, the Rx clip is placed at zero, the imaging system is placed in the range from 0.35 m to 0.46 m right in front of the QPD. The complete specifications are shown in table 2.

**Table 2.** Specifications of the four lens classical pupil plane imaging system.

|  |  | Lens 1 | Lens 2 | Lens 3 | Lens 4 | QPD |
|---|---|---|---|---|---|---|
| Radius of curvature 1 | mm | 12.009 | −8.289 | 220.27 | 40.115 |  |
| Radius of curvature 2 | mm | 376.300 | −72.758 | −16.295 | 15.609 |  |
| Position | mm | 350.000 | 365.079 | 419.745 | 456.499 | 490.0 |
| Thickness | mm | 5.500 | 2.000 | 1.5 | 3.5 |  |
| Diameter | mm | 13 | 8 | 6 | 6 |  |
| Refractive index | 1 | 1.449 63 | 1.449 63 | 1.449 63 | 1.449 63 |  |
| QPD aperture diameter | mm |  |  |  |  | 0.9 |
| QPD slit diameter | $\mu$m |  |  |  |  | 20 |

**Table 3.** Specifications of the two lens imaging system.

|  |  | Lens 1 | Lens 2 | QPD |
|---|---|---|---|---|
| Name |  | PLCX-12.7-20.6-UV-1064 | PLCC-10.0-12.9-UV-1064 |  |
| Position | mm | 350.051 | 385.109 | 434.673 |
| Primary curvature | 1 mm$^{-1}$ | 0.048 5601 | 0.0 |  |
| Secondary curvature | 1 mm$^{-1}$ | 0.0 | −0.077 4931 |  |
| Center thickness | mm | 5.3 | 2.0 |  |
| Substrate radius | mm | 6.35 | 5.0 |  |
| Refractive index | 1 | 1.449 63 | 1.449 63 |  |
| QPD aperture diameter | mm |  |  | 0.9 |
| Slit diameter | $\mu$m |  |  | 20 |
| Min. ghost power suppr. | 1 |  |  | $1.36 \cdot 10^{-7}$ |

### 5.2. Two-lens system optical design

The central idea of the two-lens imaging system is to allow a divergent exit beam for a collimated input beam and require only vanishing beam walk, vanishing tilt-to-length coupling, and a suitable beam size on the QPD. This allows a reduction in the number of lenses required while preserving the magnification factor of the beam size. The two-lens imaging systems were designed by using the framework IfoCad [13]. A list of off-the-shelf spherical fused silica lenses was used as starting point. For each combination of two lenses, the distance between both lenses as well as the distance between the second lens and QPD was varied until a measurement beam tilted by 100 $\mu$rad hit the center of the QPD at an angle of 250 $\mu$rad (0.4 magnification). For any solution found, the path length signal and its slope (tilt-to-length coupling) were computed. The solution with the best performance was chosen. The resulting set of lenses and parameters can be found in table 3.

### 5.3. Imaging system mechanical design

The mechanical hardware for the imaging systems was designed to enable high precision alignment of individual lenses. The design also allows the lenses to be given known misalignments to enable full characterization of the system's alignment sensitivity.

The design provides five degree of freedom (DoF) movement of each individual lens (no roll) and three DoF common-mode adjustment of each pair of lenses—longitudinal, transverse and yaw. The designs are customized for each of the two candidate imaging systems. All lens mounts of an imaging system are carried by a 'super-baseplate'. This super-baseplate has been designed to be adjustable. Each of the overall imaging systems has been designed to be adjustable in three DoF—longitudinal, transverse and yaw—and have the ability to be completely removed from the OB. The mechanism's compact design was driven by an optical axis of 20 mm above the top surface of the OB and by a limited footprint on the OB, avoiding mechanical interference with adjacent components. Final clamping of the aligned system is achieved through three lever clamps, whose threaded mounting base is glued to the Zerodur® OB. These lever clamps provide a downward clamping force directly to the three ball-bearing 'feet' on which the imaging system assembly sits. The complete assembly can be removed to allow the testing of different systems.

All mounts are of a thermally compensated design with the result that the central axis of the lens remains in the same position.

Adjustment of the many degrees of freedom is made using fine pitch screws and flexures, the details for which were discussed in [14]. An adjustable and removable quadrant photodiode (QPD) mount unit completes the imaging system assembly. This unit incorporates many of the design details used in the lens mounts for the imaging system with a thermally compensated design, fine vertical and horizontal QPD positioning adjustment through ultra-fine pitch screws and flexure mechanisms, and clamping of the QPD mount's kinematic baseplate to the OB baseplate with a single lever-clamp. Further design detail for the QPD mount is covered in [14].

## 6. Laser preparation and interferometer phase readout

To operate the experiment, three laser beams with different frequencies are generated on a modulation bench.

Figure 5 shows a schematic of the laser preparation and electronic setup for the heterodyne frequency generation and the phase readout. A frequency stabilized laser is split into three modulation arms where the frequency is shifted using acousto-optical modulators (AOMs). The resulting heterodyne frequencies in the interferometers are at a few kHz. Piezo mirror mounts on the modulation bench are used for an offset phase lock between the interferometer beams.

The RF frequencies driving the AOMs are generated with frequency generators that are locked to the internal clock of the phase readout system to provide stable heterodyne frequencies.

The phase readout system is a LISA Pathfinder style phasemeter with 16 input channels [15]. It is modified to be able to measure at all three heterodyne frequencies simultaneously and is also used for the offset phase locks, providing an actuation signal for the piezo mirrors on the modulation bench. Its hardware is described in [16, section C].

## 7. Construction

### 7.1. Fibre injector optical sub-assemblies

The Tx, LO and Rx Gauss beams were coupled to the optical assemblies using custom-made fibre injectors, or FIOS. The strategy was to build pre-aligned subassemblies that were then bonded to the TS and OB baseplates. The FIOS were based on the ones developed and demonstrated for LISA Pathfinder [17]. The laser beams on the TS and OB are required to travel longer distances than those on LISA Pathfinder and so the system was re-engineered to produce the larger beam waists needed to match beam size and curvature at the interference points.

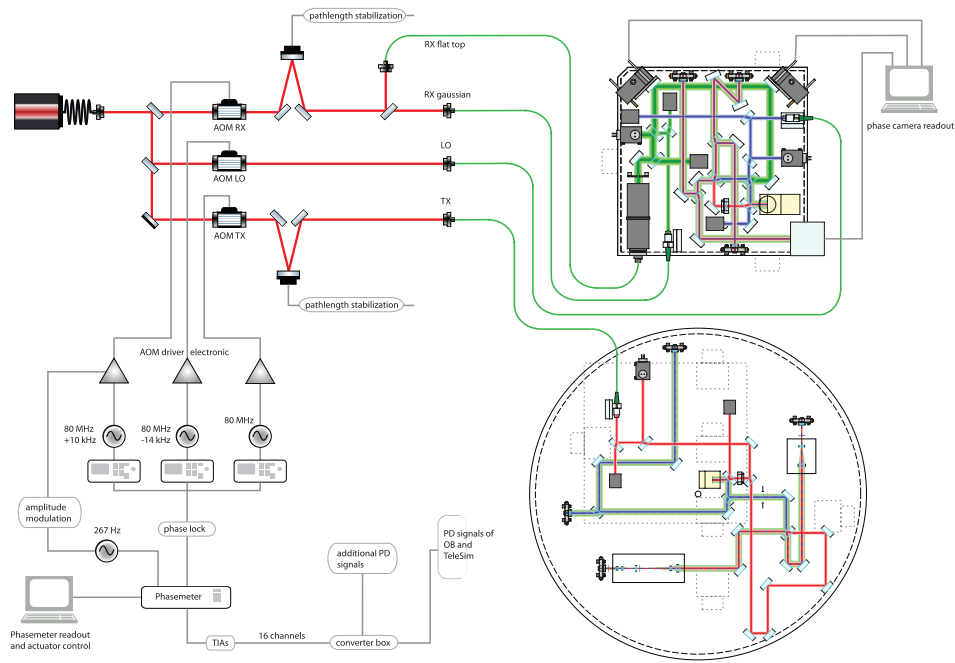

**Figure 5.** Schematic of the laser preparation and electronic setup for the heterodyne frequency generation and the phase readout. A frequency stabilized laser is split into three modulation arms where the frequency is shifted using acousto-optical modulators (AOMs), driven by RF frequency generators. Piezo mirror mounts on the modulation bench are used for an offset phase lock between the interferometer beams. The phase readout system is a modified LISA Pathfinder style phasemeter [15], which can track the three heterodyne frequencies simultaneously and provides the control signals for the offset phase locks.

A further evolution of the FIOS used here compared to the LISA Pathfinder FIOS was the use of a precision-length fused silica spacer between fibre end and collimating lens. This was implemented in an attempt to further improve the beam pointing stability with temperature, and also to allow higher laser powers to be used. In this design the power density of the laser at interfaces with air is minimised by allowing the beam to expand in glass to its collimated size, making the FIOS much less susceptible to contamination issues. This necessitated using hydroxide catalysis bonding in optical transmission with the laser propagating through up to three bonds. The beam quality achieved was excellent, with no apparent imprints on the beam intensity or wavefront by the bond layer transmissions.

The Rx beams are reflected from actuated mirrors and so their inputs do not have to be aligned to the same level of precision as the LO and Tx beams. The Rx Gauss FIOS was aligned using the technique decribed in [18]. The critical alignments of the beams from the bonded LO and Tx FIOS are the height above, and angle to, the TS and OB baseplates. The in-plane alignments are controlled by the next steering optic the beams encounter, which is aligned and bonded in place after the FIOS. The LO and Tx beams were aligned to the TS and OB baseplates respectively using the techniques described in [11, 19] to within 10 $\mu$m and 20 $\mu$rad, with measurement uncertainties of 5 $\mu$m and 20 $\mu$rad.

Figure 6 shows the Tx FIOS being bonded to the OB baseplate. Three ruby balls—two of which can be seen—are mounted on linear actuators and are used to define the position of the FIOS during bonding.

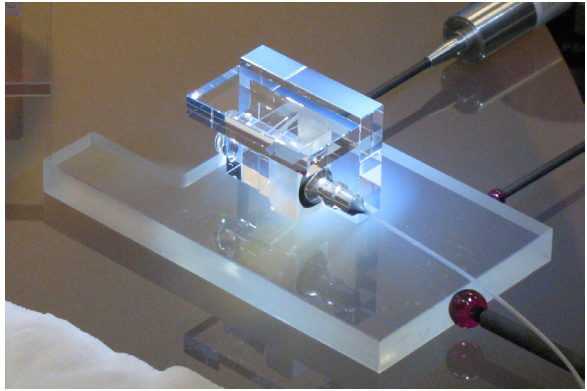

**Figure 6.** Photograph of a Tx FIOS being bonded to the OB baseplate. The FIOS has been pre-aligned in height by bonding it to a mounting post whilst monitoring and optimising the beam propagation vector relative to the L-shaped Zerodur® sub-baseplate. The ruby ball in the foreground is 8 mm in diameter.

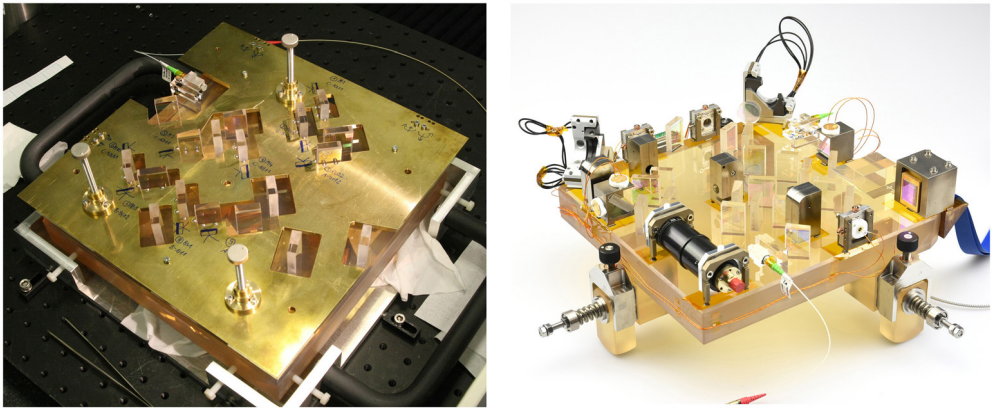

**Figure 7.** Left: Second template for bonding lower precision alignment optical components of the telescope simulator. The template has pockets with ball bearings for positioning the components to be bonded and cut outs for the already bonded components. Right: completed telescope simulator.

### 7.2. Construction of the optical bench and telescope simulator

Reflecting component alignment requirements fall in to two categories: those requiring alignment at the ∼50 $\mu$m level, and those at the sub-$\mu$m level. The lower precision alignments were achieved using CNC machined brass templates to locate the components, and the high precision alignments were achieved using the adjustable precision bonding technique—both techniques are described in [19, 20].

The precision aligned components were: PCO1 and BS11 on the TS, and BS21 on the OB. M20 and M31 on the OB were adjustably aligned to compensate for tolerance build-up. Figure 7 shows a picture of the second template used for the TS, and the completed bench with the three 'legs' of the tip-tilt mount.

All components that are not hydroxide catalysis bonded are glued to Kapton tape stuck to the baseplate. This system allows the removal of the glued components should it be required.

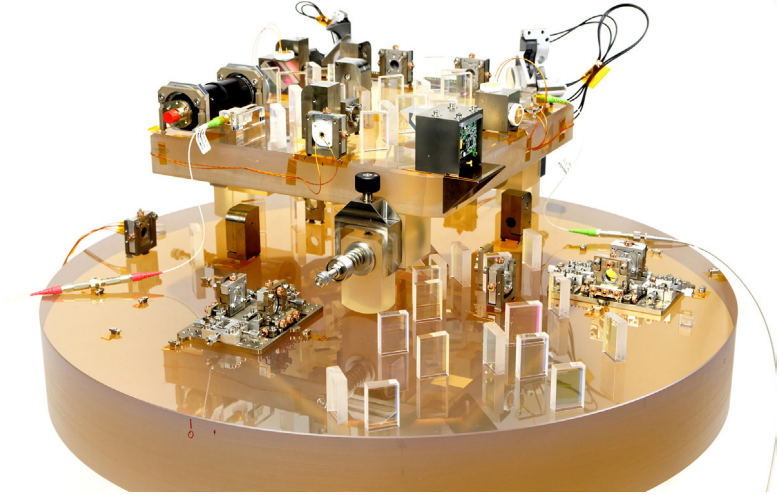

**Figure 8.** Photograph of the LISA optical bench test bed. The telescope simulator is placed on top of the optical bench.

To align the Rx flat-top beam generator and the actuators two quadrant photodiodes at different distances were used. First the Rx flat-top beam was aligned to the Rx Gauss beam, then both Rx beams to the LO beam. This alignment is not critical provided the error is within the range of the actuators.

Figure 8 shows a photograph of the LISA optical bench test bed. The telescope simulator is placed on top of the optical bench.

## 8. Tilt-to-length coupling without imaging system

Initial measurements were made without any imaging systems. Figure 9 shows the effect of beam angle changes on path length changes between science interferometer QPD1 and reference PD and their slopes. For each QPD the phase signals of their four segments are averaged (see equation (5) in [21]). From this average the phase signal of the reference single element photo detector is subtracted. The phase difference $\Delta\phi$ is converted to an optical length change $\Delta s$ according to $\Delta s = \frac{\lambda}{2\pi}\Delta\phi$ where $\lambda = 1\,064$ nm is the laser wavelength.

Three laser beams (Rx, LO, Tx) were incident on the photo detectors and led to three heterodyne signals (A: Rx–Tx, B: LO–Tx, C: Rx–LO). Both LO and Tx beams were static, while the Gaussian Rx beam was tilted around the Rx clip. Starting from angle zero the Rx beam was tilted towards negative angles, rotated back to angle zero, continued to positive angles and returned to angle zero again. Both length changes in signals A and C show a parabolic shape as is expected [6, 22]. Signal B, expected to remain constant, shows a temperature-driven drift of the telescope simulator height that will be mitigated during future measurements involving the imaging systems under test. The signals of science interferometer QPD2 are very similar to those of QPD1 and are not shown for clarity. The difference between QPD1 and QPD2 is below 6 nm for signals A and C and below 1 nm for signal B.

The graphs on the right were calculated numerically from the path length changes. For each data point, five path length data points were used in the slope calculation. The tilt-to-length coupling must not exceed $\pm 25$ $\mu$m rad$^{-1}$ for beam angles in the range $\pm 300$ $\mu$rad. Without imaging systems this requirement is violated up to a factor of six.

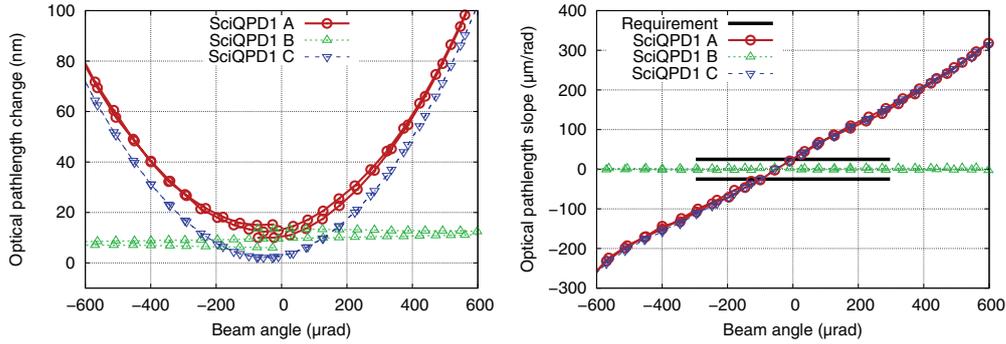

**Figure 9.** Performance measurement of the entire test bed without imaging system. On the left the pathlength change as a function of the tilt angle is shown for the three heterodyne signals (A: Rx–Tx, B: LO–Tx, C: Rx–LO) on science interferometer diode 1 (SciQPD1). Signal B is the superposition between the two static beams and therefore shows no reaction to the beam tilt. Signal A and C are the superposition between the tilting beam and one of the static beams. Both signals show the expected quadratic tilt-to-length (TTL) coupling, due to the longitudinal mismatch between point of rotation and the science interferometer detectors. The signals from the second photo detector are very similar, as intended, so only one signal is shown for clarity. In the right plot, the slope of the pathlength signal is plotted over the beam angle to relate the measurement to the path length requirement of 25 $\mu$m rad$^{-1}$, which is violated in this scenario without imaging system. In the LISA mission the requirement is relevant for signal A, the interference between Rx and Tx.

## 9. Conclusion

Tilt-to-length (TTL) coupling has a large contribution to the LISA noise budget. Suppression of this noise source is necessary and is achieved by imaging systems placed on the optical bench in front of the interferometer readout photodiodes.

We have designed and constructed an optical test bed to experimentally investigate tilt-to-length coupling. The test bed consists of a minimal optical bench and a telescope simulator. The minimal optical bench encompasses the science interferometer where the local light is interfered with light from a remote spacecraft. In our case light from the telescope simulator mimics the light from a distant spacecraft. It provides a tilting flat-top beam, a reference interferometer and an additional static beam as phase reference. We show an initial measurement of tilt-to-length coupling without imaging systems, that shows that the test bed is operational. We avoid TTL coupling in the reference interferometer by using a small photodiode placed in a copy of the beam rotation point.

We have also designed and built two different imaging systems both of which should suppress tilt-to-length coupling to the level required by LISA. We plan to report on the calibration of the test bed and the measured performance of the two imaging system designs in future publications. We also plan to perform a tolerance analysis of the two imaging systems to test their performance under the influence of intentional misalignments.

## Acknowledgments

We acknowledge funding by the European Space Agency within the project Optical Bench Development for LISA (22331/09/NL/HB), support from UK Space Agency, University of Glasgow, Scottish Universities Physics Alliance (SUPA), and support by Deutsches Zentrum für Luft und Raumfahrt (DLR) with funding from the Bundesministerium für Wirtschaft und Technologie (DLR project reference 50 OQ 0601). We thank the German Research Foundation for funding the cluster of Excellence QUEST—Centre for Quantum Engineering and Space-Time Research.